\documentclass[11pt]{article}
\usepackage{a4}
\usepackage{amsmath}
\usepackage{amssymb}
\usepackage{graphicx}
\usepackage{color}
\def\beq{\begin{equation}}
\def\eeq{\end{equation}}
\def\beqa{\begin{eqnarray}}
\def\eeqa{\end{eqnarray}}
\def\n{\nonumber \\}

\newcommand {\al}{\alpha}
\newcommand {\be}{\beta}

\begin{document}

\begin{flushright}
{SAGA-HE-270}\\
{KEK-TH-1512}
\end{flushright}
\vskip 0.5 truecm

\begin{center}
{\large{\bf 
Revisiting the Naturalness Problem
\\ -- Who is afraid of quadratic divergences? --
}}\\
\vskip 1cm

{\large Hajime Aoki$^{a}$
 and
 Satoshi Iso$^{b}$
}
\vskip 0.5cm

$^a${\it Department of Physics, Saga University, Saga 840-8502,
Japan  }\\
$^b${\it KEK Theory Center, 
High Energy Accelerator Research Organization (KEK)  \\
and 
the Graduate University for Advanced Studies (SOKENDAI), \\
Ibaraki 305-0801, Japan}
\end{center}

\vskip 1cm
\begin{center}
\begin{bf}
Abstract
\end{bf}
\end{center}
It is widely believed that quadratic divergences
severely restrict {\it natural 
constructions} of particle physics 
models beyond the standard model (SM).
Supersymmetry provides a beautiful solution, 
but the recent LHC experiments have excluded large 
parameter regions of supersymmetric extensions of the SM.
It will now be important to reconsider whether we have been
misinterpreting the quadratic divergences in field theories.
In this paper, we revisit the problem
from the viewpoint of the Wilsonian renormalization group 
and argue that quadratic divergences, which can always
be absorbed into a position of the critical surface, should
be simply subtracted in model constructions. 
Such a picture gives another justification to 
the argument \cite{FERMILAB-CONF-95-391-T}
that the scale invariance of the SM, 
except for the soft-breaking terms,
 is an alternative solution
to the naturalness problem.
It also largely broadens possibilities of model constructions
beyond the SM since
we just need to take care of 
 logarithmic divergences, which cause
mixings of various physical scales 
and runnings of couplings.

\newpage
\section{Introduction}
\label{sec:introduction}   
\setcounter{footnote}{0}
\setcounter{equation}{0}

The hierarchy problem \cite{145528}, the stability
of the weak scale against Planck or GUT scales, 
is considered to be an important guiding principle
to construct a model beyond the standard model (SM).
Supersymmetry is a
beautiful solution, but the recent LHC experiments have already 
excluded low energy supersymmetry and we need to 
solve the little hierarchy problem
as well as the $\mu$ problem. 
Faced with these difficulties,
it will  be  important and timely 
to reconsider the 
hierarchy problem and ask 
 whether we have been
misinterpreting the divergences in field theories.

The hierarchy problem has many faces and it is important 
to distinguish the following two types.
The first one, which is most commonly referred,
is why a scalar field can be much lighter than 
the cutoff scale.\footnote{The meaning of a cutoff is twofold. 
It is either a cutoff in a UV complete theory
or a cutoff in an effective theory at which new degrees of freedom appear. 
We use it in both meanings  and explain their differences 
depending on the situations.}
The mass of a scalar field receives large radiative corrections 
due to quadratic divergences,
and the tree-level mass of the Higgs field and the loop contributions 
of the cutoff order must cancel to a very high precision in order of the weak scale. 
Another type of the hierarchy problem, which
is caused by logarithmic divergences,
arises when a theory includes multiple physical scales, 
e.g., the weak scale and the GUT scale 
\cite{Print-76-0529 (HARVARD)}. 
Even if the quadratic divergence is disposed of,
the lower  mass scale generically
receives  large radiative corrections of the higher mass scales,
and a fine-tuning is necessary to keep the separation of  the multiple  scales.

The first type of the hierarchy problem
can be regarded as an academic question since a subtraction of 
the quadratic divergences is 
always possible without any physical effect on low energy dynamics.\footnote{
In a UV complete theory, quadratic divergence is only an artifact of a
regularization procedure while, in an effective theory, 
it is interpreted as a boundary condition
between a UV complete and an effective theory. In both cases,
quadratic divergences can be subtracted without any physical effect on low
energy dynamics.}
It is  quite different from the logarithmic divergences, which play  important
physical roles as beta functions or conformal anomalies.  
In some formulations,
one can sidestep the quadratic divergences and accordingly  
the associated hierarchy problem.
A well-known example is to use the dimensional regularization \cite{74886}.
Another way may be  just to subtract them \cite{arXiv:1104.3396}. 
A crucial feature in those formulations is that one can 
separate the subtractive and multiplicative 
renormalization procedures.
Then the first type of the hierarchy problem is reduced to the naturalness
of such a subtraction.

As argued in ref.~\cite{FERMILAB-CONF-95-391-T},
scale invariance can be used to justify a subtracted theory,
analogously to imposing the supersymmetry. 
At the classical level, the SM is scale invariant except for the Higgs mass term,
and a vanishing of the mass term would increase the symmetry of the SM. 
The common wisdom is that such increase of the symmetry cannot play any role to
control the divergences because scale invariance is broken by  
logarithmic runnings of coupling constants.
However, quadratic divergences are independent of the logarithmic divergences
and  merely an artifact of regularization procedures, 
and hence should be simply subtracted.
Then the trace of the energy-momentum tensor becomes
\beq
 \Theta^\mu_\mu 
 = \Delta m^2 H^{\dagger} H + \beta_{\lambda_i} {\cal O}_i \ ,
\eeq
where  $\Delta m^2$ is {\it not} proportional 
to the cutoff squared $\Lambda^2$ but to  $m^2$.
The quadratic divergences are subtracted 
and the mass term gets multiplicatively renormalized.
Here, the anomalous term as well as the mass term is regarded as soft-breaking terms
of the scale invariance
since they do not generate quadratic divergences.

In this paper,
we first argue that 
a subtraction of quadratic divergences   
is naturally performed 
from the Wilsonian renormalization group (RG) \cite{81239}
point of view,
and give another justification for a subtracted theory.
In the Wilsonian RG,
 quadratic divergences 
determine a position of the critical surface 
in the theory space, and 
the scaling behavior of RG flows around the critical surface 
is determined only by the logarithmic divergences. 
The subtraction of the quadratic divergences can thus be
performed by the position of the critical surface.
Then, the subtraction is interpreted as a choice of
the parameterization, 
i.e., a coordinate transformation, in the theory space.
The fine-tuning we need to perform  
is the distance of bare parameters from the critical surface, 
and has nothing to do with the position of the critical surface itself.
Hence, quadratic divergences are not the real issue of
the fine-tuning problem \cite{wetterich, grange}.

If we consider an effective low energy theory with a finite cutoff,
we encounter quadratic divergences of the cutoff order.
When we embed the effective theory in a UV complete fundamental theory, however,
they are compensated by the same kind of divergences
arising from the integrations above the cutoff of the effective theory.
The subtraction in the effective theory is then justified by
the boundary condition at the cutoff.
Once the boundary condition is determined by the dynamics 
in the UV complete fundamental theory, 
the quadratic divergences that appear in the effective theory
can be legitimately subtracted by the boundary condition.
One can also use the above arguments of coordinate transformation in the theory space
to justify the subtraction within the effective theory.

The second type of the hierarchy problem 
is more physical and should be taken with much care.
In the Wilsonian RG, it is formulated as
a radiative mixing  of  multiple relevant operators,
which is caused by logarithmic divergences.
The lower mass scale is  affected by higher scales through RG transformations.
The mixing is physical and of course cannot
be simply subtracted.
Hence it gives a strong constraint on 
{\it natural} model building.

The above arguments broaden possibilities of model constructions
beyond the SM.  
In particular,  nonsupersymmetric 
models with quadratic divergences
but no large logarithmic mixings can be good
candidates of models beyond the SM.
Examples of such models are
$\nu$MSM \cite{Asaka} or a classically conformal
TeV scale  $B-L$ model \cite{IOO}.

The paper is organized as follows. 
We first discuss quadratic divergences
in the Wilsonian RG in section \ref{sec:RGflow},
by studying a $\phi^4$ field theory at the 
one-loop order.
We then argue that 
the quadratic divergence is naturally subtracted
and is not the real issue of the 
fine-tuning problem in section \ref{sec:finetune}.
We further discuss logarithmic divergences 
and the second type of the hierarchy
problem in section \ref{sec:2scales}.
Finally in section \ref{sec:HO}, 
we show that
our statements hold at all orders in perturbations.
The last section is devoted to conclusions and discussions.

\section{RG flows of $\phi^4$ theory at one-loop}
\label{sec:RGflow}
\setcounter{equation}{0}

In this section we explain the role of quadratic divergences
in the Wilsonian renormalization group (RG).
As an example, we consider
 a scalar field theory 
in d=4 at the one-loop approximation.
The theory has both of the quadratic and logarithmic divergences, but 
the quadratic divergences can be completely absorbed 
into the position of the critical surface.
In section \ref{sec:HO} we see that it holds generally 
at all orders in perturbation expansions.

We first consider a single scalar theory with a $\phi^4$  interaction 
on a $d$-dimensional Euclidean lattice.
Its action is given by
\beqa
S&=&\int_{\Lambda^d} \frac{d^d p}{(2\pi)^d} \frac{1}{2}(p^2+m^2)\phi(p)\phi(-p) \n
&&+\frac{1}{4!}\lambda \int_{\Lambda^d} \prod_{a=1}^{4} \frac{d^d p_a}{(2\pi)^d}
(2\pi)^d \delta^{(d)}(\sum_{a=1}^{4} p_a)~\phi(p_1)\phi(p_2)\phi(p_3)\phi(p_4)  \ ,
\label{actionphi4}
\eeqa
where the momentum integration is performed over the region
\beq
\Lambda^d = \{p|-\pi< p^i < \pi, \ \forall i=1,2,\cdots,d \} \ .
\label{Lambdad}
\eeq
All the quantities in the action, i.e.,
the parameters $m^2$ and $\lambda$, the scalar field $\phi$, and the cutoff $\Lambda=\pi$,
are dimensionless.

An RG transformation can be defined by the following two steps. \\
{\it Step 1: Integration over higher momentum modes}  \\
For simplicity, we introduce a sharp boundary at $p=\pi/N$ with a constant $N>1$
and divide the integration region $\Lambda^d$ 
into two regions with lower and higher momenta,
\beqa
\Lambda^d_{\rm in} &=& 
\{p|-\frac{\pi}{N}< p^i < \frac{\pi}{N}, \ \forall i=1,2,\cdots,d \} \ , \\
\Lambda^d_{\rm out} &=& 
\{p| \ |p^i| \ge \frac{\pi}{N}, \ \exists i=1,2,\cdots,d \} \ .
\label{momregion}
\eeqa
We then perform functional integrations over 
$\phi(p)$ with $p \in \Lambda^d_{\rm out}$.
The remaining theory is described by an effective action 
for the lower momentum modes,
 $\phi(p)$ with $p \in \Lambda^d_{\rm in}$. \\
{\it Step 2: Rescalings}  \\
We then rescale the momentum $p$ and the field $\phi(p)$ as
\beqa
p' &=& N p \ , \n
\phi'(p') &=& N^{-\theta} \phi (p) \ .
\eeqa
This rescaling of momenta makes the integration region back to
the original one, $\Lambda^d$. 
The scaling dimension $\theta$ can be chosen so that 
the scalar field has the canonical kinetic term.
$(-\theta)$ is the mass dimension of the field $\phi(p)$ and
 given by the canonical value $\theta = (d+2)/2$ 
near the Gaussian fixed point.\footnote{
If we consider RG flows near a nontrivial fixed point, 
$\theta$ deviates from the canonical value
due to an anomalous dimension of the field.
Near the Gaussian fixed point,
we can keep $\theta$ to be the canonical one and
absorb all effects of the wave function renormalization
in the coefficients $c_n$ given below. 
At one-loop of the $\phi^4$ theory, a wave function renormalization
is  absent.}

In the d=4 $\phi^4$ theory, we can restrict RG transformations
 within the space of two parameters $m^2$ and $\lambda$.
The other operators, such as $\phi^6$ or $p^4\phi^2$,
become less and less important when repeating the RG transformations and
eventually become negligible: they are irrelevant operators.
Hence, the restriction to consider the RG transformations
within the subspace is justified.
The resulting theory  has the same form as the original one in (\ref{actionphi4}),
but with the parameters changed. 

\begin{figure}
\begin{center}
\includegraphics[height=3cm]{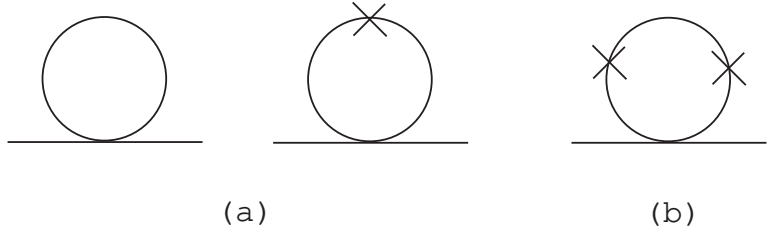}
 \caption{(a) Feynman diagrams that contribute to the mass renormalization 
 transformation (\ref{RGtr1mass}). The cross represents a mass insertion.
 (b) A diagram in higher order in $m^2$, which does not contribute to (\ref{RGtr1mass})
in the limit (\ref{molll}).}
\label{fig:masstr}
\end{center}
\end{figure}

\begin{figure}
\begin{center}
\includegraphics[height=3cm]{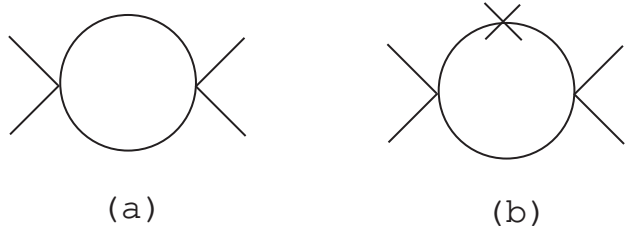}
 \caption{(a) A Feynman diagram that contributes to the coupling renormalization 
 transformation (\ref{RGtr1lambda}).
 (b) A diagram in higher order in $m^2$, which does not contribute to (\ref{RGtr1lambda})
 in the limit (\ref{molll}).}
\label{fig:couplingtr}
\end{center}
\end{figure}

We perform the above functional integrations over 
$\phi(p)$ with $p \in \Lambda^d_{\rm out}$
by perturbative expansions with respect to the coupling $\lambda$.
At the one-loop order, 
the above two steps give the following changes of the parameters: 
\beqa
m^{2 \ \prime} &=& 
N^{2\theta-d} (m^2+c_1 \lambda -c_2 m^2 \lambda) \ , 
\label{RGtr1mass}\\
\lambda' &=&  
N^{4\theta-3d} (\lambda -3c_2 \lambda^2) \ , 
\label{RGtr1lambda}
\eeqa
where $c_1$ and $c_2$ are positive constants, 
\beqa
c_1 &=& \frac{1}{2} \int_{\Lambda^d_{\rm out}} 
\frac{d^dq}{(2\pi)^d} \frac{1}{q^2} >0 \ , \label{c1} \\
c_2 &=& \frac{1}{2} \int_{\Lambda^d_{\rm out}} 
\frac{d^dq}{(2\pi)^d}
\left(\frac{1}{q^2}\right)^2 >0 \ . \label{c2}
\eeqa
The prefactors $N^{2\theta-d}$ and $N^{4\theta-3d}$ come from the rescalings (Step 2),
while $c_n$ are contributions from the integrations of higher momentum modes (Step 1).
The mass transformation (\ref{RGtr1mass}) is given 
by the diagrams in Figure \ref{fig:masstr}(a).
Since the integration is performed in the UV region $\Lambda^d_{\rm out}$
and no IR divergences occur, 
we can expand the propagator with respect to $m^2$.
In eq.~(\ref{RGtr1mass}) we  took the first two terms in the expansion.
Higher order terms in $m^2$,
such as $c'_2 m^4 \lambda$ which comes from a diagram in Figure \ref{fig:masstr}(b), 
are highly suppressed  since we suppose that 
\beq
\frac{m^2}{\Lambda^2} \ll 1 \ .
\label{molll}
\eeq
Similarly  the coupling transformation (\ref{RGtr1lambda}) at one-loop is determined
 by the diagram in Figure \ref{fig:couplingtr}(a).
Again, higher order terms in $m^2$,
such as $c'_2 m^2 \lambda^2$ for Figure \ref{fig:couplingtr}(b), 
do not contribute to (\ref{RGtr1lambda}) due to the assumption (\ref{molll}).
There are no wave function renormalizations at the one-loop order.
The dependence of the constants $c_n$ 
on the cutoff $\Lambda$ is easily derived.\footnote{In taking the continuum limit, 
we consider a limit where the physical momentum becomes infinitesimal compared to the 
cutoff $\Lambda=\pi$. Hence, $\Lambda=\pi$ corresponds to a large momentum scale 
in the unit of the physical scale.} 
By evaluating the integrals of (\ref{c1}) and (\ref{c2}), we find
\beqa
c_1 &\propto& \Lambda^{d-2} (1-N^{-(d-2)})~\longrightarrow~ 
\Lambda^{2} (1-N^{-2})~~{\rm for}~d=4 \ , \\
c_2 &\propto& \Lambda^{d-4} (1-N^{-(d-4)})~\longrightarrow~ 
\Lambda^{0} \ln N~~{\rm for}~d=4 \ .
\eeqa
Hence, $c_1$ and $c_2$ reflect the quadratic and the 
logarithmic divergences, respectively.

By performing the RG transformations (\ref{RGtr1mass}) 
and (\ref{RGtr1lambda}) several times,
one obtains RG flows in the theory space, 
the space spanned by the parameters $\lambda$ and $m^2$ 
in this case. 
Since eq.~(\ref{RGtr1lambda}) depends only on $\lambda$, but not on $m^2$,
the flow of the coupling constant $\lambda$ is determined only by using (\ref{RGtr1lambda}).
For $d \neq 4$,
the equation (\ref{RGtr1lambda}) is rewritten as
\beq
\frac{1}{\lambda'}-\frac{1}{\lambda^*} =  
N^{-(4\theta-3d)} \left(\frac{1}{\lambda}-\frac{1}{\lambda^*}\right) +{\cal O}(\lambda) \ , 
\ \ \ \ 
\lambda^* = \frac{N^{4\theta-3d}-1}{3c_2} \ .
\eeq
After performing the transformation $n$ times, one obtains
\beq
\frac{1}{\lambda_n}-\frac{1}{\lambda^*} =  
N^{-(4\theta-3d)n} \left(\frac{1}{\lambda_0}-\frac{1}{\lambda^*}\right) \ .
\label{flowlambda}
\eeq
Here, $\lambda_n$ is the renormalized coupling constant after $n$-times RG transformations.
There are two fixed points of the RG transformation: one with $\lambda=0$, 
and the other with $\lambda=\lambda^*$.
For $d=4$, two fixed points coincide at $\lambda=0$  and one finds  
\beq
\frac{1}{\lambda_n} =\frac{1}{\lambda_0} +3 c_2 n   \ .
\label{flowlambda4d}
\eeq
If $\lambda_0 >0$, $\lambda_n$ approaches the fixed point $\lambda=0$
as one increases $n$.

In order to obtain the flow in the direction of  $m^2$, we rewrite eq.~(\ref{RGtr1mass}), 
by using (\ref{RGtr1lambda}), as
\beq
m^{2 \ \prime} - m^2_c(\lambda') 
= N^{2\theta-d} (1-c_2 \lambda)(m^2 - m^2_c(\lambda) ) \ ,
\label{RGtr1massre}
\eeq
with a function $m^2_c(\lambda)$, determined up to this order as
\beq
m^2_c(\lambda) = -\frac{c_1}{1- N^{2(\theta-d)}} \lambda \ .
\label{critline}
\eeq
Performing the RG transformation $n$ times, one obtains
\beqa
m^{2}_n - m^2_c(\lambda_n) 
&=& N^{(2\theta-d)n} \prod_{i=0}^{n-1} (1-c_2 \lambda_i)~(m^2_0 - m^2_c(\lambda_0) ) \n
&\simeq& N^{(2\theta-d)n} \exp(-c_2 \sum_{i=0}^{n-1} \lambda_i)~
(m^2_0 - m^2_c(\lambda_0) ) \ .
\label{flowm}
\eeqa
The equation $m^2=m_c^2(\lambda)$ determines the position of the critical line,
and eq.~(\ref{flowm}) shows how the {\it distance} from the critical line 
scales under the RG transformations.

\begin{figure}
\begin{center}
\includegraphics[height=7cm]{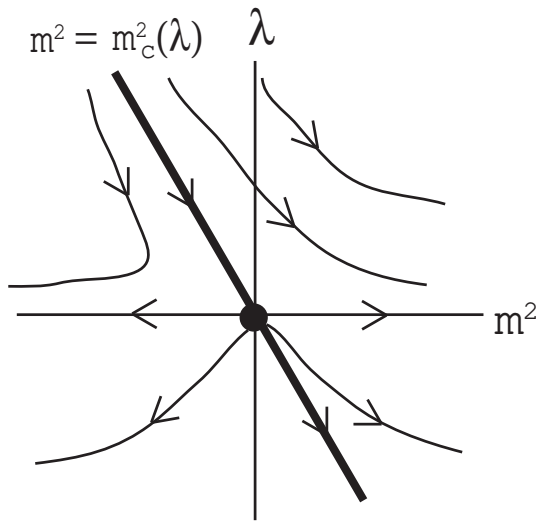}
 \caption{RG flow for $d=4$. The blob indicates the Gaussian fixed point,
 and the thick line $m^2=m^2_c(\lambda)$ corresponds to the critical line.}
\label{fig:RGflow}
\end{center}
\end{figure}

The RG flow is given by the two equations (\ref{flowm}) and (\ref{flowlambda}),
or (\ref{flowlambda4d}) for $d=4$.
A schematic picture of the flow for the $d=4$ case is drawn in Figure \ref{fig:RGflow}.
The fixed point of the RG flow is given by $m^2=\lambda=0$,
i.e., the Gaussian fixed point.
If the initial parameters are exactly on the critical line $m^2=m^2_c(\lambda)$, 
they continue to be there after RG transformations and approach  the fixed point. 
We consider only the region $\lambda \ge 0$
for the stability of vacuum.
A theory on the critical line is a massless theory.
If the initial parameters are off the critical line,
they depart from it with $\lambda$ decreasing under the RG transformations.

As we stressed in the introduction, the constant $c_1$,
which reflects the quadratic divergences,
is completely absorbed into the position of the critical line, 
i.e., the definition of the function $m^2_c(\lambda)$ in (\ref{critline}),
and the scaling behavior (\ref{flowm}) of the RG flow  around the critical line 
is determined only by $c_2$, which reflects the logarithmic divergences.  
It is a very important message
we can read from the Wilsonian treatment of the 
renormalization group.
As we see in the next section,
it corresponds to the fact that
in ordinary perturbative calculations,
quadratic divergences can always be subtracted
without any physical effects on the dynamics of 
field theories, unlike logarithmic divergences.

\section{Quadratic divergences, subtractions, and the fine-tuning problem}
\label{sec:finetune}
\setcounter{equation}{0}

We now interpret the subtraction and
 the fine-tuning problem in the framework of the Wilsonian RG,
and argue that the quadratic divergence can be naturally subtracted  
and is not the 
real issue of the hierarchy problem. 

\subsection{Continuum limit}
We first review how to take the continuum limit.
In the constructive formulation of a field theory,
one usually takes the continuum limit by simultaneously letting the parameters
close to the critical surface and taking the cutoff to infinity.
For a scalar theory in $d=4$, however, one cannot construct an interacting theory
in this way, which is well-known as the triviality or the Landau singularity problem.
We thus consider a theory at a large but finite cutoff with bare parameters 
very close to the critical surface.

We first write the RG equation (\ref{flowm})  as
\beq
m^{2}_{0} - m^2_c(\lambda_{0}) 
= N^{-(2\theta-d)n} e^{c_2 \sum_{i=0}^{n-1} \lambda_{i}}~(m^2_n - m^2_c(\lambda_n) ) \ .
\label{flowmback}
\eeq
The parameters $m^2_0$ and $\lambda_0$ on the left-hand side (LHS) describe the bare parameters,
while $m^2_n$ and $\lambda_n$ on the right-hand side (RHS) describe the renormalized ones.
Note that all the quantities in this relation are dimensionless, including
the cutoff $\Lambda=\pi$.
We now introduce dimensionful cutoffs.
First we introduce the low energy scale $\tilde{\Lambda}_n=M$, e.g., $M=100$ GeV.
The renormalized parameters $m_n$ and $\lambda_n$  are defined at this scale.
The higher scale  where the bare parameters $m_0$, $\lambda_0$ 
are defined is given by
\beq
\tilde\Lambda_{0}=N^{n}\tilde\Lambda_n = N^n M \ .
\label{ratio_lambda0n}
\eeq

Let us express everything in terms of dimensionful physical quantities instead of
the dimensionless lattice parameters.
Physical momentum $\tilde{p}$ is defined by
\beq
\tilde p_k = \frac{\tilde{\Lambda}_k}{\Lambda} p \ , \ \ \
\tilde{\Lambda}_k = N^{n-k} M \ , \
\eeq
for each cutoff theory at $k=0,1,\cdots,n$. 
We can similarly define the field as
$\tilde{\phi}_k(\tilde{p}_k)=(\tilde{\Lambda}_k/\Lambda)^{-\frac{d+2}{2}} \phi(p) $.
The dimensionless parameters $m^2,\lambda$ are also replaced by 
a dimensionful mass and a coupling 
\beq
\tilde m^2_k = \left(\frac{\tilde{\Lambda}_k}{\Lambda}\right)^{2} m^2 \ , \ \ \ 
\tilde \lambda_k = \left(\frac{\tilde{\Lambda}_k}{\Lambda}\right)^{4-d}\lambda \ .
\label{tildemm}
\eeq
In terms of these physical quantities, the RG equation (\ref{flowmback}) can be rewritten as
\beq
\tilde m^{2}_{0} - \tilde m^2_c(\lambda_{0}) 
= N^{(2-2\theta+d)n}e^{c_2 \sum_{i=0}^{n-1} 
\lambda_{i}}~(\tilde m^2_n - \tilde m^2_c(\lambda_n) ) \ .
\label{flowmbackdimful}
\eeq
The critical value $m^2_c(\lambda)$ is also rescaled as
\beqa
\tilde m^{2}_c(\lambda_k) 
= \left(\frac{\tilde \Lambda_k}{\Lambda}\right)^2 m^{2}_c(\lambda_k) \ ,
\label{tildemcmc}
\eeqa
which reflects quadratic divergences of the position of the critical line,
as shown in the previous section.
For the Gaussian fixed point, we have $\theta=(d+2)/2$,
and hence $N^{(2-2\theta+d)n}=1$. 
Eq.~(\ref{flowmbackdimful}) becomes
\beq
(\tilde m^{2}_{0}-\tilde m^2_c(\lambda_{0}) ) =
e^{c_2 \sum_{i=0}^{n-1} \lambda_{i}}~(\tilde m^2_n -\tilde m^2_c(\lambda_n)) \ . 
\label{massrelation}
\eeq
The prefactor $e^{c_2 \sum_{i=0}^{n-1} \lambda_{i}}$  represents
the logarithmic running of the mass parameter. 
Indeed, from (\ref{flowlambda4d}), one can find $\lambda_i \sim 1/3c_2 i$,
and thus the prefactor behaves as $e^{c_2 \sum_{i=0}^{n-1} \lambda_{i}} 
\sim n^{1/3} \sim (\ln \tilde\Lambda_0)^{1/3}$.
Its $n$-dependence is weaker than a power-law running behavior 
$N^{xn} \sim (\tilde\Lambda_0)^x$ with some constant $x$.

In ordinary perturbative calculations of field theories, 
we first choose a regularization method  and a renormalization prescription
to deal with divergences.
Quadratic divergences are simply subtracted with appropriate renormalization conditions.
The so called fine-tuning problem is that we need to fine-tune
the bare mass $\tilde{m}_0^2$ of a scalar field at the cutoff scale $\tilde{\Lambda}_0$, 
so that the renormalized mass  $\tilde{m}_R^2$, 
\beqa
 \tilde{m}_R^2 
 &=& \tilde{m}_0^2 
 + \alpha(\lambda, \ln(\tilde{\Lambda}_0/\tilde{\Lambda}_n))~\tilde{\Lambda}_0^2  
 + \beta(\lambda, \ln(\tilde{\Lambda}_0/\tilde{\Lambda}_n))~\tilde{m}_0^2 \n
 &=& (1+\beta(\lambda, \ln(\tilde{\Lambda}_0/\tilde{\Lambda}_n)))
 ~(\tilde{m}_0^2 + \alpha'(\lambda, \ln(\tilde{\Lambda}_0/\tilde{\Lambda}_n))~\tilde{\Lambda}_0^2 ) \ ,
\label{masstuning} 
\eeqa
is held at the desired value.
Here $\alpha$ and $\beta$ are functions of the coupling $\lambda$
and $\ln (\tilde{\Lambda}_0/\tilde{\Lambda}_n)$,
and assumed  to be
independent of the mass parameter $\tilde m_0$.  
Namely,
the  mass-independent renormalization scheme \cite{'tHooft:1973mm, Weinberg:1951ss}
is chosen.
When we apply (\ref{masstuning}) to the Higgs mass,
this looks very unnatural because, 
if $\tilde{\Lambda}_0 \gg \tilde{m}_R=m_{\rm W}$, the tree-level  mass $\tilde{m}_0^2$
of the Higgs particle and the loop contributions $\alpha\tilde{\Lambda}_0^2$ 
must cancel to a very high precision in order of the weak scale $m_{\rm W}^2$. 
There is also a subtlety in 
the mass-independent renormalization 
for theories with quadratic divergences \cite{Weinberg:1951ss, arXiv:1104.3396}.

Let us make a comparison between eq.~(\ref{massrelation}) 
and eq.~(\ref{masstuning}).
The bare mass parameter $\tilde m^2_0$ and the quadratic divergence 
$\alpha' \tilde{\Lambda}_0^2$ in the RHS of (\ref{masstuning}) 
correspond to the bare mass $\tilde m^2_0$ and the critical value $\tilde m^2_c(\lambda_0)$ 
in the LHS of (\ref{massrelation}), respectively.
The renormalized mass $\tilde m_R^2$ of  (\ref{masstuning})
is given by
$\tilde m^2_n -\tilde m^2_c(\lambda_n)$ in the RHS of (\ref{massrelation}).
The multiplicative renormalization factor $(1+ \beta)$ in (\ref{masstuning})
corresponds to
the factor for the logarithmic running
 $e^{-c_2 \sum_{i=0}^{n-1} \lambda_{i}}$ in (\ref{massrelation}).

\subsection{Subtractive renormalization in Wilsonian RG}

In the Wilsonian RG, 
the quadratic divergence $c_1$ is absorbed
into the position of the critical line $\tilde m^2_c(\lambda)$. 
Note also that all observable quantities like the correlation length
are determined by the distance of the mass parameter $\tilde m_0^2$ from the 
critical line $\tilde m^2_c(\lambda_0)$, not by the value $\tilde m_0^2$ itself. 
The fact that
the difference $\tilde m_0^2-\tilde m^2_c(\lambda_0)$ 
is the physically relevant quantity
gives a natural interpretation for
the subtractive renormalization of the quadratic divergences.

As we stressed in the introduction,
an important and necessary feature for the  subtracted theories,
as in \cite{74886, arXiv:1104.3396, FERMILAB-CONF-95-391-T},
where one can sidestep the quadratic divergences,
is that one can separate
the subtractive and the multiplicative renormalization procedures.
Looking at eq.~(\ref{massrelation}),
the prefactor $e^{c_2 \sum_{i=0}^{n-1} \lambda_{i}}$ 
corresponds to the multiplicative renormalization.
Hence, eq.~(\ref{massrelation}) clearly 
separates the subtractive,
$\tilde m^{2}_{0}-\tilde m^2_c(\lambda_{0})$,
and the multiplicative,
$e^{c_2 \sum_{i=0}^{n-1} \lambda_{i}}$,
renormalization procedures.

Another remarkable result from eq.~(\ref{massrelation})
is as follows.
Neither the prefactor $e^{c_2 \sum_{i=0}^{n-1} \lambda_{i}}$ 
nor the critical value $\tilde m^2_c(\lambda)$  depends on
the mass parameter $\tilde m_0$.
Hence,
eq.~(\ref{massrelation})  gives an explicit realization of 
the mass-independent renormalization scheme \cite{'tHooft:1973mm, Weinberg:1951ss}.
As discussed in \cite{Weinberg:1951ss, arXiv:1104.3396},
whether or not one can formulate a mass-independent renormalization scheme
in theories with quadratic divergences
was a nontrivial and subtle problem.

Note that 
the subtraction is not arbitrary in the Wilsonian RG.
Once we choose a scheme to calculate the RG flow, the critical line
is given unambiguously.
Seeming ambiguities only come from our lack of calculability
of the exact RG flows, but RG flows are in principle exactly 
determined once we choose a scheme. 
The only fine-tuning we need is the {\it distance} 
from the critical surface to the bare parameters.

Before discussing the tuning of the distance, 
let us convince ourselves that the position of the critical surface has nothing
to do with the fine-tuning problem 
by comparing two theories: one with a nonzero $c_1$, and the other with a vanishing 
$c_1$. The latter theory corresponds to a theory with vanishing quadratic divergences.
The critical lines are given by
$m_c^2 \ne 0$ for $c_1 \ne 0$, and $m_c^2 = 0$ for $c_1 = 0$.
Is the first theory  more unnatural than the second one?
The feeling of  unnaturalness simply comes from  seeming ambiguities of 
calculating the position of the critical line, but as stressed, once a scheme is given
it is determined unambiguously and there are no distinctions between 
these two theories.
In other words, we can take  new coordinates of the theory space such that the critical 
line is parallel to an axis of the coordinates. Then a theory with $c_1 \ne 0$ looks the same 
as a theory with $c_1=0$. The change of coordinates can be determined exactly
once we choose a scheme.  In the dimensional regularization, quadratic divergences do not
appear. We may say that this corresponds to taking a new coordinate 
by $m_{\rm new}^2 = m^2  - m_c^2(\lambda)$. 
Then the critical line is given by $m_{\text{new},c}^2(\lambda)=0$
even for a theory with $c_1 \ne 0.$
Therefore
the subtraction of the quadratic divergences
is simply a choice of the coordinates of the theory space.
In this sense, the quadratic divergences are 
naturally subtracted and are
not the real issue of the hierarchy problem.

Finally, we consider the tuning of the distance from the critical line,
namely, the fine-tuning to choose the bare mass   
near the critical value with the precision of the weak scale. 
Such a fine-tuning is kinematical in the following sense. 
Looking at eq.~(\ref{flowmback}), we see that 
this kind of  tuning comes from the factor $N^{-(2\theta-d)n}=N^{-2n}$
for the Gaussian fixed point,
but it reflects
 nothing but the canonical dimension of the mass and 
has nothing to do with the  quadratic divergences. 
This fine-tuning is necessary as long as the dimension of mass square is two.
The prefactor $e^{c_2 \sum \lambda_i}$ 
simply gives a logarithmic scaling factor $n^{1/3}$, and 
does not change the situation much.

If we consider a nontrivial fixed point instead of the Gaussian one,
the canonical scaling dimension is modified by anomalous dimensions of the mass and the field.
For example, if  $\theta =(d+2)/2 -\delta$ and the coupling constant at the fixed point is given by
$\lambda_*$, the relation (\ref{massrelation}) is modified as\footnote{
The prefactor $e^{c_2 \sum_{i=0}^{n-1} \lambda_{i}}$ in (\ref{massrelation}) 
gives not only the factor $e^{c_2 \lambda_* n}$ in (\ref{massrelation2}), but also
an extra factor $e^{c_2 \sum_{i=0}^{n-1} (\lambda_{i}-\lambda_*)}$. 
Since it has a weaker $n$-dependence than a power-law behavior 
$N^{xn} \sim \tilde\Lambda^x$, we neglected it here.
The wave function renormalization also gives a deviation from $\theta$, $\delta$
and an extra factor  neglected in (\ref{massrelation2}).}
\beq
(\tilde m^{2}_{0}-\tilde m^2_c(\lambda_{0}) ) \sim 
(N^{2\delta } e^{c_2 \lambda_* } )^n
(\tilde m^2_n -\tilde m^2_c(\lambda_n) ) \ .
\label{massrelation2}
\eeq
Thus, large anomalous dimensions can relax the condition of
how precisely we need to define the bare mass near the critical line.
If one feels uncomfortable with the relation (\ref{massrelation}), a possible resolution
will be to construct a theory with large anomalous dimensions.

\section{Mixing of  multiple scales}
\label{sec:2scales}
\setcounter{equation}{0}

In the previous two sections, we considered a theory with a single physical
scale besides the cutoff scale. 
In this section, we study an RG flow in a theory with  hierarchically separated  multiple scales.
These  scales are mixed by radiative corrections
associated with  the logarithmic divergences. 
This causes the second type of the hierarchy problem.
We emphasize that
it again has nothing to do with the quadratic divergences.

For simplicity, we consider a theory with multiple scalar fields 
on a $d$-dimensional Euclidean lattice,
whose action is given by
\beq
S=\int_{\Lambda^d} \left[ \sum_{\alpha =1}^{S} \left(\frac{1}{2}(p^2+m^2_\alpha)\phi_\alpha^2
+\frac{1}{4!}\lambda_{\alpha\alpha} \phi_\alpha^4\right)
+\sum_{\alpha\ne\beta} \frac{1}{8}\lambda_{\alpha\beta}
\phi_\alpha^2\phi_\beta^2 \right] \ ,
\eeq
with $\lambda_{\alpha\beta} = \lambda_{\be\al}$.
The index $\alpha=1,\ldots,S$ labels species of the scalar fields.
There are $S$ mass parameters $m^2_\alpha$
and $S(S+1)/2$ coupling parameters $\lambda_{\alpha\beta}$ in this case.
The momentum integration region $\Lambda^d$ is taken as in eq.~(\ref{Lambdad}).

To obtain an RG transformation, we follow the two steps
 in section \ref{sec:RGflow}.
At the one-loop order, it becomes
\beqa
m^{2 \ \prime}_\alpha &=& N^{2\theta-d} 
\left[m^2_\alpha+c_1 \sum_\beta\lambda_{\alpha\beta} 
-c_2 \sum_{\beta} \lambda_{\alpha\beta} m^2_\beta  \right] \ , \label{RGtr1massa} \\
\lambda'_{\alpha\alpha} &=&  N^{4\theta-3d} \left[\lambda_{\alpha\alpha} 
-3c_2 \sum_{\beta}(\lambda_{\alpha\beta})^2\right] \ , \label{RGtr1lambdaaa} \\
\lambda'_{\alpha\beta} &=&  N^{4\theta-3d} \left[\lambda_{\alpha\beta} 
-c_2 \sum_{\gamma} \lambda_{\alpha\gamma}\lambda_{\beta\gamma}
-4c_2 (\lambda_{\alpha\beta})^2\right] \ , \label{RGtr1lambdaab}
\eeqa
with $c_1$ and $c_2$ given in (\ref{c1}) and (\ref{c2}).

By performing the RG transformations several times,
one obtains RG flows in the theory space, 
 $S(S+3)/2$-dimensional space spanned by the parameters 
$\lambda_{\alpha\beta}$ and $m^2_\alpha$.
We use $\lambda$ to represent the set of couplings $\lambda_{\alpha \beta}$ in the following.
Since eqs. (\ref{RGtr1lambdaaa}) and (\ref{RGtr1lambdaab})
depend only on $\lambda_{\alpha\beta}$, and not on $m^2_\alpha$,
the flow in $\lambda_{\alpha\beta}$ is determined by 
(\ref{RGtr1lambdaaa}) and (\ref{RGtr1lambdaab}).
For the flow in $m^2_\alpha$, we can rewrite eq.~(\ref{RGtr1massa}) as
\beq
m^{2 \ \prime}_\alpha - m^2_{c\alpha}(\lambda') 
= N^{2\theta-d} \sum_\beta
(\delta_{\alpha\beta}-c_2 \lambda_{\alpha\beta})
(m^2_\beta - m^2_{c\beta}(\lambda) ) \ ,
\label{RGtr1massare}
\eeq
where $m^2_{c\alpha}(\lambda)$ are functions of $\lambda_{\alpha\beta}$
and are determined by the equations
\beq
m^2_{c\alpha}(\lambda') - N^{2\theta-d} \sum_\beta
(\delta_{\alpha\beta}-c_2 \lambda_{\alpha\beta}) m^2_{c\beta}(\lambda)
= N^{2\theta-d}  c_1 \sum_\beta\lambda_{\alpha\beta} \ ,
\eeq
together with (\ref{RGtr1lambdaaa}) and (\ref{RGtr1lambdaab}).
The solutions are given as a power series of $\lambda$ as
\beq
m^2_{c\alpha}(\lambda) = -\frac{c_1}{1- N^{2(\theta-d)}} 
\sum_{\beta}\lambda_{\alpha\beta} 
+ {\cal O}(\lambda^2) \ .
\label{critsurfa}
\eeq
 The quadratic divergence in d=4, namely $c_1$, is again absorbed 
into the position of the critical surface $m_\alpha^2 = m^2_{c \alpha}(\lambda)$.

Now let us define a symmetric mixing matrix
\beq
(M_{(k)})_{\alpha \beta} =
 \delta_{\alpha \beta}- c_2 \lambda_{\alpha \beta (k)}
\simeq \exp (\delta_{\alpha \beta}- c_2 \lambda_{\alpha \beta (k)}) \ ,
\eeq
and their product
\beq
M = M_{(n-1)} M_{(n-2)} \cdots M_{(0)} \ .
\eeq
Here $\lambda_{\alpha \beta (k)}$ represents the coupling constant $\lambda_{\alpha \beta}$
after $k$-times RG transformation.
By performing the transformation (\ref{RGtr1massare})  $n$ times, one obtains
\beq
m^{2}_{\alpha (n)} - m^2_{c\alpha}(\lambda_{(n)}) 
= N^{(2\theta-d)n}~\sum_{\beta} M_{\alpha\beta}
~(m^2_{\beta (0)} - m^2_{c\beta}(\lambda_{(0)}) ) \ .
\label{flowma}
\eeq

In taking the cutoff very large, 
one needs to fine-tune the  $S$ relevant operators, $m_{\alpha(0)}^2$. 
We rewrite the RG equation (\ref{flowma}) 
as 
\beq
m^{2}_{\alpha (0)} - m^2_{c\alpha}(\lambda_{(0)}) 
= N^{-(2\theta-d)n}\sum_{\beta}
(M^{-1})_{\alpha\beta}
(m^2_{\beta (n)} - m^2_{c\beta}(\lambda_{(n)}) ) \ ,
\label{mixRG}
\eeq
where
\beq
(M^{-1})_{\alpha\beta} = \delta_{\alpha\beta} + c_2 
\sum_{k=0}^{n-1} \lambda_{\alpha\beta (k)}
+ {\cal O}(\lambda^2) \ . 
\label{M-1c2mixing}
\eeq
The equation  (\ref{mixRG}) can be written in terms of the dimensionful parameters, 
as in (\ref{massrelation}), as
\beq
\tilde m^{2}_{\alpha (0)} - \tilde m^2_{c\alpha}(\lambda_{(0)}) 
=  \sum_{\beta}\left(M^{-1}\right)_{\alpha\beta}
~(\tilde m^2_{\beta (n)} - \tilde m^2_{c\beta}(\lambda_{(n)}) ) \ .
\label{flowmadim}
\eeq
Here we used the canonical value of $\theta=(d+2)/2$ for the Gaussian fixed point.

We now simplify the discussion by considering a theory with two separate renormalized 
scales, e.g.,
\beqa
\tilde m^2_{1 (n)} - \tilde m^2_{c1}(\lambda_{(n)}) &=& m_{\rm W}^2 \ , \n
\tilde m^2_{2 (n)} - \tilde m^2_{c2}(\lambda_{(n)}) &=& m_{\rm GUT}^2 \ .
\eeqa
It then follows from (\ref{flowmadim}) that 
we need  to fine-tune the bare mass parameters such that the difference of 
$(\tilde m^2_{1,0} - \tilde m^2_{c1}(\lambda_{0}) )$ and radiative corrections
from the higher physical scale $m^2_{\text{GUT}}$ are canceled to give
the weak scale:
\beq
m_{\rm W}^2 
\simeq  
\frac{1}{(M^{-1})_{11}}
(\tilde m^2_{1,0} - \tilde m^2_{c1}(\lambda_{0}) ) -
 \frac{(M^{-1})_{12}}{(M^{-1})_{11}} ~m_{\rm GUT}^2 
\ .
\label{2ndHPfinetune}
\eeq
Unlike the subtraction of the critical mass parameter $m_c^2(\lambda)$,
the GUT scale $m^2_{\rm GUT}$ is a physically observable scale 
and cannot be subtracted. 
Or, in other words, the second term proportional to $m^2_{\rm GUT}$ cannot
be absorbed into the position of the critical surface.
Hence, unless the off-diagonal element of the matrix $(M^{-1})_{12}$
is suppressed, we need a fine-tuning of the bare mass 
$(\tilde m^2_{1,0} - \tilde m^2_{c1}(\lambda_{0}))$
against the GUT scale with a high precision in order of the weak scale. 
As we can see from (\ref{M-1c2mixing}),
in order to solve the problem, we need
to suppress either the coefficient $c_2$,
the mutual couplings $\lambda_{\alpha\beta}$,
or the higher scale $m^2_{\rm GUT}$.

\section{Higher orders of perturbations}
\label{sec:HO}
\setcounter{equation}{0}

Up to now, our discussions are based on the one-loop order calculations,
but our statements hold at all orders of  perturbations in the coupling $\lambda$.
In this section, we extend the statement in section \ref{sec:RGflow}  
to all orders. 
It is also straightforward to extend the results in the other sections in the same way.

In the following, by using a
 renormalized perturbation theory, we will show iteratively 
that the statement in section 2 does hold 
at all orders of perturbations 
in the coupling constant $\lambda$.
We first replace the mass parameter $m^2$ by $m^2-m_c^2+m_c^2$ in the action
(\ref{actionphi4}),
\beq
S=\int_{\Lambda} \left[ 
\frac{1}{2}p^2\phi^2+\frac{1}{2}(m^2-m_c^2(\lambda))\phi^2 
+\frac{1}{2} m_c^2(\lambda) \phi^2
+\frac{1}{4!}\lambda \phi^4 \right]  \ ,
\label{actionren}
\eeq
where the $\lambda$ dependence of the 
position of the critical surface $m_c^2(\lambda)$  will be determined later
in a self-consistent way.
We then perform the functional integrations over the higher momentum modes
(Step 1 of the RG transformation)
by perturbative expansions with respect to $m^2-m_c^2(\lambda)$ and $m_c^2(\lambda)$,
as well as $\lambda$.
Such mass expansions are legitimate
 since the integrations are performed only in the
UV region and free from IR divergences.
An insertion of the renormalized mass parameter $m^2-m_c^2(\lambda)$ 
suppresses loop integrations
by a factor $1/\Lambda^2$, 
and thus always appears in a combination of 
\beq
\frac{m^2-m_c^2(\lambda)}{\Lambda^2} 
= {\cal O}\left(\frac{1}{\Lambda^2}\right) \ .
\label{m2mc2Lam-2}
\eeq
Here, we assumed that $m^2-m_c^2(\lambda)$ is of order $\Lambda^0$.
It was explicitly 
shown at the one-loop order in sections \ref{sec:RGflow},
and will be justified later in this section 
in an iterative way at higher orders in $\lambda$.
We can thus neglect higher order terms in the expansion of $m^2-m_c^2(\lambda)$.
On the other hand, an insertion of the critical mass
 is not suppressed by $1/\Lambda^2$
since the value $m_c^2(\lambda)$ itself is of order
$\Lambda^2$:
\beq
\frac{m_c^2(\lambda)}{\Lambda^2} 
= {\cal O}\left(\lambda\right) \ .
\label{mc2lam}
\eeq
Though it is suppressed by the coupling constant
$\lambda$, it is not suppressed by the cutoff $\Lambda$.
It was already shown in (\ref{critline}) at one-loop order,
and can be justified later in this section in an iterative way.
Hence, on the contrary to (\ref{m2mc2Lam-2}), 
we  need to 
take higher orders of $m_c^2(\lambda)$ in the 
perturbative calculations.

\begin{figure}
\begin{center}
\includegraphics[height=2.4cm]{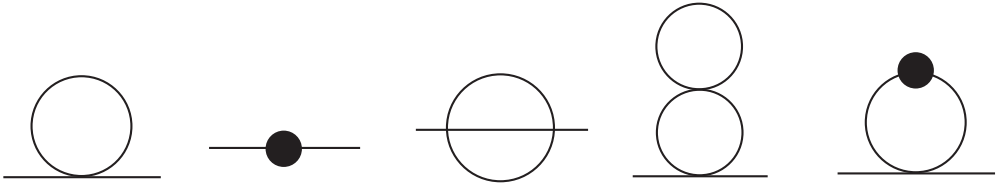}
 \caption{The Feynman diagrams that contribute to $f\Lambda^2$ up to order $\lambda^2$.
 The dot represents an insertion of $m_c^2$.
 The first two diagrams give order $\lambda^1$ contributions, 
 while the last four give order $\lambda^2$.
 Note that the second diagram gives both order $\lambda^1$ and $\lambda^2$ contributions.}
\label{fig:f12}
\end{center}
\end{figure}

\begin{figure}
\begin{center}
\includegraphics[height=4.5cm]{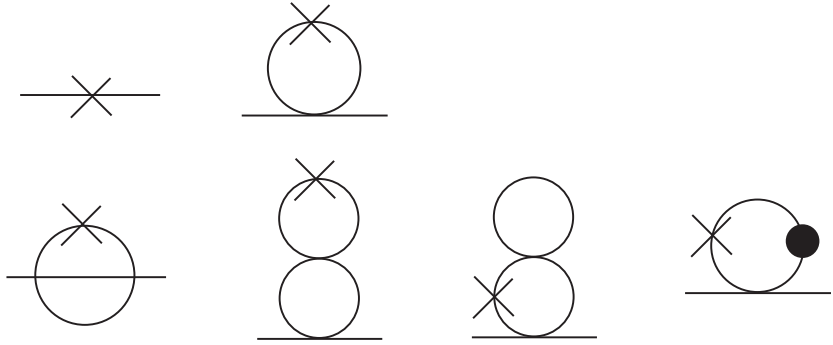}
 \caption{The Feynman diagrams that contribute to $g$ up to order $\lambda^2$.
 The cross and the dot represent an insertion of $m^2-m_c^2$ and $m_c^2$, respectively.
 The first diagram gives an order $\lambda^0$ contribution, 
 the second $\lambda^1$,
 and the last four $\lambda^2$.}
\label{fig:g012}
\end{center}
\end{figure}

\begin{figure}
\begin{center}
\includegraphics[height=4.2cm]{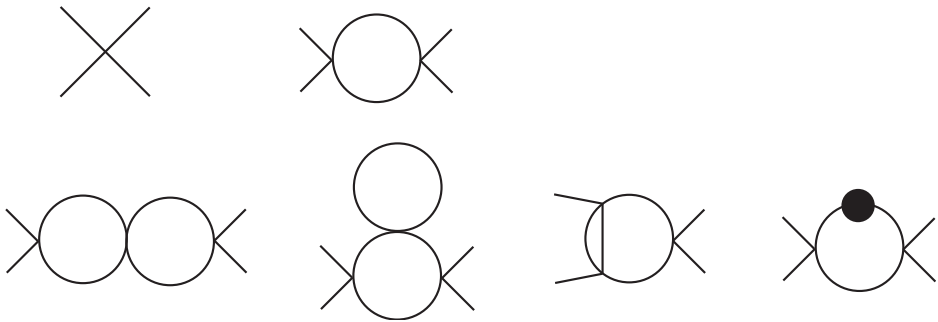}
 \caption{The Feynman diagrams that contribute to $h$ up to order $\lambda^3$.
 The dot represents an insertion of $m_c^2$.
 The first diagram gives an order $\lambda^1$ contribution, 
 the second $\lambda^2$, 
 and the last four $\lambda^3$.}
\label{fig:h123}
\end{center}
\end{figure}

As a result of the above arguments, we can schematically 
write the effective action after the
integrations as
\beqa
&&\int_{\Lambda/N} \Bigg[ 
\frac{1}{2}~e(\lambda,\frac{m_c^2}{\Lambda^2})~p^2\phi^2
+\frac{1}{2}\left[f(\lambda,\frac{m_c^2}{\Lambda^2})~\Lambda^2 
+ g(\lambda,\frac{m_c^2}{\Lambda^2})~(m^2-m_c^2(\lambda))\right]\phi^2 \n
&&+\frac{1}{4!}~h(\lambda,\frac{m_c^2}{\Lambda^2})~\phi^4 \Bigg]  \ ,
\label{actionstep1}
\eeqa
where the functions $e$, $f$, $g$ and $h$ are power series with respect to
$\lambda$ and $m_c^2/\Lambda^2$, with the coefficients dependent on $N$.
$f\Lambda^2$ is given by the 2-point, 1PI diagrams with the external momentum fixed to be zero.
They are depicted in Figure \ref{fig:f12} up to order $\lambda^2$,
where the dot represents an insertion of $m_c^2$.
$g$ is given by those diagrams with a single insertion of $m^2-m_c^2(\lambda)$.
They are depicted in Figure \ref{fig:g012} up to order $\lambda^2$,
where the cross represents an insertion of $m^2-m_c^2$.
$e-1$ is given by the diagrams of $f$, but the first term in the expansion
of the external momentum.
They begin with the third diagram in Figure \ref{fig:f12} at order $\lambda^2$. 
Similarly, $h$ is given by the 4-point, 1PI diagrams with the external momentum fixed to be zero.
They are depicted in Figure \ref{fig:h123} up to order $\lambda^3$,

We next perform the wave function renormalization,
$\phi \to e^{-1/2} \phi$,
and the rescaling of momenta and fields.
It is Step 2 of the RG transformation, 
and we have 
\beq
\int_{\Lambda} \left[ 
\frac{1}{2}p^2\phi^2
+\frac{1}{2} N^{2\theta-d} e^{-1}\left[f~\Lambda^2 
+ g~(m^2-m_c^2(\lambda))\right]\phi^2 \n
+\frac{1}{4!} N^{4\theta-3d }e^{-2}h~\phi^4 \right]  \ .
\label{actionstep2}
\eeq
We then find a generalized 
formula of  (\ref{RGtr1mass}) and (\ref{RGtr1lambda}), as 
given by the equations
\beqa
m^{2 \ \prime} 
&=& N^{2\theta-d} e^{-1}
\left[f~\Lambda^2 
+ g~(m^2-m_c^2(\lambda))\right] \ , 
\label{RGtrmassfull} \\
\lambda' 
&=&  N^{4\theta-3d} e(\lambda,\frac{m_c^2(\lambda)}{\Lambda^2})^{-2}~
h(\lambda,\frac{m_c^2(\lambda)}{\Lambda^2})
\ . \label{RGtrlambdafull}
\eeqa
As in eq.~(\ref{RGtr1massre}), we rewrite (\ref{RGtrmassfull}) as
\beq
m^{2 \ \prime} -m_c^2(\lambda')
= N^{2\theta-d} e(\lambda,\frac{m_c^2(\lambda)}{\Lambda^2})^{-1} 
g(\lambda,\frac{m_c^2(\lambda)}{\Lambda^2})~(m^2-m_c^2(\lambda)) \ , 
\label{massscalingfull}
\eeq
with 
\beq
m_c^2(\lambda')
= N^{2\theta-d} e(\lambda,\frac{m_c^2(\lambda)}{\Lambda^2})^{-1} 
f(\lambda,\frac{m_c^2(\lambda)}{\Lambda^2})~\Lambda^2 \ .
\label{criticallinefull}
\eeq
The functional form of $m_c^2(\lambda)$ is determined  iteratively in $\lambda$ by solving 
the equation (\ref{criticallinefull}), with (\ref{RGtrlambdafull}) 
inserted in the LHS of (\ref{criticallinefull}).
At the order of $\lambda^1$, $f$ is given by the first two diagrams in Figure \ref{fig:f12}.
Then, the solution $m_c^2(\lambda)$ indeed becomes (\ref{critline}).
At the order of $\lambda^2$, $f$ is given by the last four diagrams in Figure \ref{fig:f12}.
The dot in the loop of the last diagram means the order $\lambda^1$ term in $m_c^2(\lambda)$,
while the dot in the second diagram means the order $\lambda^2$ term in $m_c^2(\lambda)$.
$e$ has the order $\lambda^2$ term, but it does not affect 
(\ref{criticallinefull}) at order $\lambda^2$.
The LHS of (\ref{criticallinefull}) has both contributions 
from the order $\lambda'^1$ term in $m_c^2(\lambda')$ 
with the $\lambda^2$ term of (\ref{RGtrlambdafull}),
and from the order $\lambda'^2$ term in $m_c^2(\lambda')$.
In this way, we could obtain the $\lambda^2$ term in $m_c^2(\lambda)$.
We can also obtain higher order terms iteratively in the same way.

The most important property of  eq.~(\ref{criticallinefull}) is that 
the solution $m^2_c(\lambda)$ is proportional to $\Lambda^2$ at all orders in perturbations.
Following the same arguments given in section \ref{sec:RGflow},
we can see that $m^2_c(\lambda)$ gives the position of the critical line,
and (\ref{RGtrlambdafull}) and (\ref{massscalingfull}) 
determine the scaling behavior of the RG flows
around the critical line.
We therefore find that at all orders in perturbative expansions in $\lambda$,
the quadratic divergences are completely absorbed in the position of 
the critical line, and they do not play any role in
the dynamics of field theories. The RG flow around the critical line is determined only
by the logarithmic divergences.
Hence, the assumptions (\ref{m2mc2Lam-2}) and (\ref{mc2lam}) we adopted in this section
are justified iteratively in $\lambda$.

As we mentioned in section \ref{sec:finetune}, eq.~(\ref{massscalingfull})
means that
we can separate the subtractive and multiplicative renormalization procedures.
Eqs. (\ref{RGtrlambdafull}) and (\ref{massscalingfull}) also give an explicit
representation of the mass-independent renormalization scheme.
The separability of quadratic and logarithmic divergences
relies on several properties of loop integrations 
in Wilsonian RG.
First of all, they are free from IR divergences
and hence, we can perform the mass expansions.
Secondly, the loop integrations are performed from
$\Lambda/N$ to $\Lambda$,
and thus a divergence from the subdiagram is compensated by
the remaining part of the diagram with negative power of $\Lambda$.
For instance, in the 5th diagram of Figure \ref{fig:g012},
while the upper subdiagram gives a $\Lambda^2$ contribution, 
the lower one gives $\Lambda^{-2}$, which makes 
the overall diagram of order $\Lambda^0$ and insensitive to the divergence from the subdiagram.
We thus need to take care of only the overall superficial degrees of divergences\footnote{
Our renormalized perturbation is reminiscent of the usual one of BPHZ.
In our case, however, the cancellation between the loop contribution and the counter term 
is not exact.
For example, at order $\lambda^1$, the first and the second diagram in 
Figure \ref{fig:f12} did not cancel in (\ref{criticallinefull}).
It follows that, at order $\lambda^2$,
the loop contribution of the upper subdiagram in the 4th diagram 
is not canceled by the dot in the 5th diagram.
One can also see the same situation in the last two diagrams in Figure \ref{fig:g012}.
However,
divergences from subdiagrams cause no problem in our case.
}.
That is why the naive dimensional analysis is possible.

\section{Conclusions and discussions}
\label{sec:conclusion}
\setcounter{equation}{0}

In this paper, we revisited the hierarchy problem, i.e., stability 
of  mass of a scalar field against large radiative corrections,
from the Wilsonian RG point of view.
We first saw that  quadratic divergences can be
absorbed into a position of the critical surface $m_c^2(\lambda)$,
and the scaling behavior of RG flows around the critical surface is 
determined only by  logarithmic divergences.
The subtraction of the quadratic divergences
is unambiguously fixed by the critical surface. 
In another word, the subtraction is interpreted as 
taking a new coordinate of the space of parameters such that
$m_{\rm new}^2=m^2-m_c^2(\lambda)$. 
These arguments gave a natural interpretation for the subtractions,
and another justification for the subtracted theories
as in \cite{74886, arXiv:1104.3396, FERMILAB-CONF-95-391-T}.
The fine-tuning problem, i.e., the hierarchy between the physical scalar mass
and the cutoff scale,
is then reduced to a problem of taking the bare mass parameter close 
to the critical surface in taking the continuum limit.
It has nothing to do with the quadratic divergences in the theory.
Therefore the quadratic divergences are not the real issue of the hierarchy problem. 
If we are considering a low energy effective theory with an effective cutoff,
the subtraction of the quadratic divergences corresponds to taking 
a boundary condition at the effective cutoff scale.
Hence it has nothing to do with the dynamics at a lower energy scale,
and when such divergences appear in radiative corrections,
we can simply subtract them.

We also considered another type of the hierarchy problem.
If a theory consists of  multiple physical scales, e.g., the weak scale $m_{\rm W}$ and 
the GUT scale $m_{\rm GUT}$ 
besides the cutoff scale $\tilde{\Lambda}$, 
a mass of the lower scale $m_{\rm W}$ receives large radiative corrections 
$\delta m_{\rm W}^2 \propto m_{\rm GUT}^2 \log \tilde{\Lambda}/16\pi^2$ through
the logarithmic divergences.
Such a mixing of physical mass scales is interpreted as a mixing of relevant
operators along the RG flows. 
Unlike the first type of the hierarchy problem, the mass of the larger scale $m_{\rm GUT}$
 cannot simply be disposed of  by a subtraction.
In order to solve such a mixing problem, we  need  to suppress the mixing by 
some additional conditions. Of course, if these two scalars are extremely weakly
coupled, the mixing can be suppressed. If the couplings are not so weak,
we need to cancel the mixing by symmetries or some nontrivial dynamics.
A well-known example is the supersymmetry, 
where the non-renormalization theorem assures the absence of such mixings
unless supersymmetry is broken.

Let us comment on scheme dependence of the subtraction. 
In this paper, we fixed one scheme to perform  RG transformations.
Then the critical surface is unambiguously determined.
If we change the scheme, e.g., from the sharp cut off (\ref{momregion}) of the higher mode 
integrations to another one \cite{Polchinski:1983gv}, 
the RG transformations are changed, and so accordingly is the position
of the critical surface. 
But the definition of bare parameters is correlated with the choice of  the scheme.
Hence a shift of the position of the critical surface by a change of  scheme
does not mean an ambiguity of the critical surface. 
Rather it corresponds to changing coordinates of the theory space. 
In this sense, the subtraction of a position of the critical surface 
from the bare mass is performed for each fixed scheme without any ambiguity.

What is the meaning of  the coefficients of various terms in the bare action?
In the investigation of the RG flows, we encountered two kinds of quantities,
the mass parameter $\tilde{m}^2$ and the subtracted mass $(\tilde{m}^2-\tilde{m}_c(\lambda)^2)$.
The issue of the fine-tuning problem, i.e., the stability against the quadratic divergences, 
is related to which quantity we should consider to be a physical parameter.
In the renormalization-group-improved field theory, as we studied in this paper, the subtracted one
is considered to be physical. 
The mass parameter itself depends 
on a choice of coordinates of the theory space, and changes scheme by scheme.
If we want to {\it derive} low energy field theories from a more fundamental theory,
like a string theory, the bare parameter itself, $\tilde{m}^2$, is widely
believed to be related to the fundamental quantities at the string scale.
But since such a quantity is coordinate (of the theory space) dependent, 
we should use the subtracted one
when we relate low energy  field theories with more fundamental theories. 
The amount of subtraction is given by the boundary condition
at the cutoff, and determined by the dynamics at higher scales
in the fundamental theory. It is independent of the low energy
dynamics.

We are thus left with the second type of the hierarchy problem, namely
a mixing of the weak scale with another physical scale like $m_{\rm GUT}.$
We classify possible ways out of it:
\begin{enumerate}
\item SM up to $\tilde\Lambda$ 
\item New physics around TeV, but nothing beyond up to $\tilde\Lambda$ 
\item New physics at a higher scale, but extremely weakly coupled with SM
\item New physics at a higher scale with nontrivial dynamics or symmetries
\end{enumerate}
The first possibility is to consider a model without any further physical scale 
up to the cutoff scale $\tilde\Lambda$. 
The Planck scale may play a role of a cutoff scale for the SM.
As we saw in this paper, the quadratic divergence of 
the cutoff order can be simply subtracted
and it does not cause any physical effect. In the second possibility, 
we introduce a new scale
which may be coupled with the SM, but  suppose that  
the new scale is not so large compared with the weak scale. 
Then even if the mixing is not so small, the weak scale does not 
receive large radiative corrections. 
Various kinds of TeV scale models 
are classified into this category.
Some examples are
$\nu$MSM \cite{Asaka}
and the classically 
conformal\footnote{Classical conformality means absence of mass terms for scalar fields.
The model of course receives radiative corrections, but
classical conformality is not broken by logarithmic divergences.}
 TeV-scale $B-L$ extended model \cite{IOO}. 
The third one is to consider a very large physical scale, but
with the mutual coupling suppressed to be very small.
The final possibilities include  a supersymmetric GUT, but the 
low energy theory of  broken supersymmetry must be supplemented 
with the second type of scenario. 
If we worry about quadratic divergences,
the first three categories need fine-tunings against the cutoff scale
and are excluded by the naturalness condition. 
Hence, most model building beyond the SM has been restricted to the last category.
Once we admit that quadratic divergence is not the real issue of the
hierarchy problem, it broadens our possibilities of 
model constructions. 

The discussions of the quadratic divergences in this paper
can also be extended to the quartic divergences, 
and if gravity is described in terms of a renormalized field theory, 
the Wilsonian RG treatment might give a new perspective of 
the cosmological constant problem. 
We hope to come back to this issue in future. 


\end{document}